\documentclass[aps, prx,noshowpacs,reprint,superscriptaddress]{revtex4-1}
\usepackage{amsmath}
\usepackage{pifont}
\usepackage{amssymb}
\usepackage{tipa}
\usepackage{amsfonts}
\usepackage{mathrsfs}
\usepackage{graphicx}
\usepackage{dcolumn}
\usepackage{bm}
\usepackage{color}
\usepackage{booktabs}
\usepackage[colorlinks,linkcolor=red,citecolor=blue]{hyperref}
\usepackage{threeparttable}
\usepackage{multirow}
\usepackage{epstopdf}
\begin{document}

\title{Multiferroic van der Waals heterostructure FeCl$_2$/Sc$_2$CO$_2$: \\Nonvolatile electrically switchable electronic and spintronic properties}

\author{Liemao Cao}
\affiliation{College of Physics and Electronic Engineering, Hengyang Normal University, Hengyang 421002, China}

\author{Xiaohui Deng}
\affiliation{College of Physics and Electronic Engineering, Hengyang Normal University, Hengyang 421002, China}

\author{Guanghui Zhou}
\affiliation{Department of Physics, Key Laboratory for Low-Dimensional Structures and Quantum Manipulation (Ministry of Education), Hunan Normal University, Changsha 410081, China}
\affiliation{Deparment of Physics, College of Sciences, Shaoyang University, Shaoyang 422001, China}

\author{Shi-Jun Liang}
\affiliation{National Laboratory of Solid State Microstructures, School of Physics, Collaborative Innovation Center of Advanced Microstructures, Nanjing University, Nanjing 210093, China}

\author{Chuong V. Nguyen}
\affiliation{Department of Materials Science and Engineering, Le Quy Don Technical University, Ha Noi 100000, Viet Nam}

\author{L. K. Ang}
\affiliation{Science, Mathematics and Technology, Singapore University of Technology and Design, 8 Somapah Road, Singapore 487372}

\author{Yee Sin Ang}
\email{yeesin\_ang@sutd.edu.sg}
\affiliation{Science, Mathematics and Technology, Singapore University of Technology and Design, 8 Somapah Road, Singapore 487372}

\begin{abstract}
Multiferroic van der Waals (vdW) heterostrucutres offers an exciting route towards novel nanoelectronics and spintronics device technology.
Here we investigate the electronic and transport properties of multiferroic vdW heterostructure composed of ferromagnetic FeCl$_2$ monolayer and ferroelectric Sc$_2$CO$_2$ monolayer using first-principles density functional theory and quantum transport simulations.
We show that FeCl$_2$/Sc$_2$CO$_2$ heterostructure can be reversibly switched from semiconducting to half-metallic behavior by electrically modulating the ferroelectric polarization states of Sc$_2$CO$_2$.
Intriguingly, the half-metallic phase exhibits a Type-III broken gap band alignment, which can be beneficial for tunnelling field-effect transistor application.
We perform a quantum transport simulation, based on a \emph{proof-of-concept} two-terminal nanodevice, to demonstrate all-electric-controlled valving effects uniquely enabled by the nonvolatile ferroelectric switching of the heterostructure.
These findings unravels the potential of FeCl$_2$/Sc$_2$CO$_2$ vdW heterostructures as a building block for designing a next generation of ultimately compact information processing, data storage and spintronics devices.
\end{abstract}

\maketitle


\section{Introduction}
The continuous miniaturization of electronic products posses huge technological challenges on the design of nanoelectronic devices and functional materials.
Spintronics and two-dimensional (2D) materials provides new ideas for solving this problem.
Compare to the conventional spintronics based on three-dimensional (3D) bulk magnetic materials, 2D-material-based spintronics offers enormous opportunities for the post Moore's law era due to its low energy consumption, fast device operation, mechanical flexibility, high storage density, and gate-controllable operation \cite{Soumyanarayanan,Zhang}.
The recent advancements of van der Waals (vdW) heterostructure, obtained via the vertical stacking of different 2D materials, further expands the design flexibility and the functionality of 2D-material-based nanoelectronic, optoelectronics and spintronics devices \cite{liemao,li,liang,liu,Gong1,zhong,Novoselov, ANG_PRB, ANG_PRB2, Cao_APL, Qing_APL, Chuong_PRB, Chuong_NJC, Chuong_JPCL, bafekry2020a,bafekry2020b,bafekry2020c,bafekry2020d}.

2D materials with intrinsic magnetism are uncommon and often required complex nanofabrication techniques \cite{Shi, Tongay, jiazhang}.
Nonetheless, the recent theoretical \cite{Guo} and experimental \cite{Gong, Huang} advancements in 2D materials with intrinsic magnetism have revitalized the exploration of intrinsic magnetism in 2D layered materials for spintronic applications\cite{zhong,song}, such as the very large negative magnetoresistance ratio of over 10,000\% in the vdW tunnel junction made of ultrathin magnetic semiconductor CrI$_3$ \cite{Kim,Wang}, 100\% spin polarization in tunnel junction composed of CrI$_3$ tunnel barrier \cite{Paudel}, valley splitting and valley-polarized electroluminescence in ultrathin CrI$_3$ vdW heterostructures \cite{Zhang1,Seyler}.
Beyond 2D magnetic materials, \emph{2D ferroelectric monolayer} has also received intensive research attention in recent years \cite{Cui}, primarily because of their huge potential in memory and neuromorphic devices \cite{Guan} as enabled by the presence of electrically-switchable spontaneous ferroelectric polarization.
The recent exploration of intrinsic ferroelectric properties in 2D monolayers, such as In$_2$Se$_3$ \cite{Ding}, 1T-phase MoS$_2$ \cite{Shirodkar}, MX (M=Ge, Sn; X=S, Se, Te) \cite{Fei1,Wang2}, bilayer BN \cite{Li1}, and CuInP$_2$S$_2$ \cite{Qi}, have shed much light on the physics and device application potential of 2D ferroelectric materials.
\begin{figure*}[t]
\centering
\includegraphics[bb=0 35 800 580, width=6.0 in]{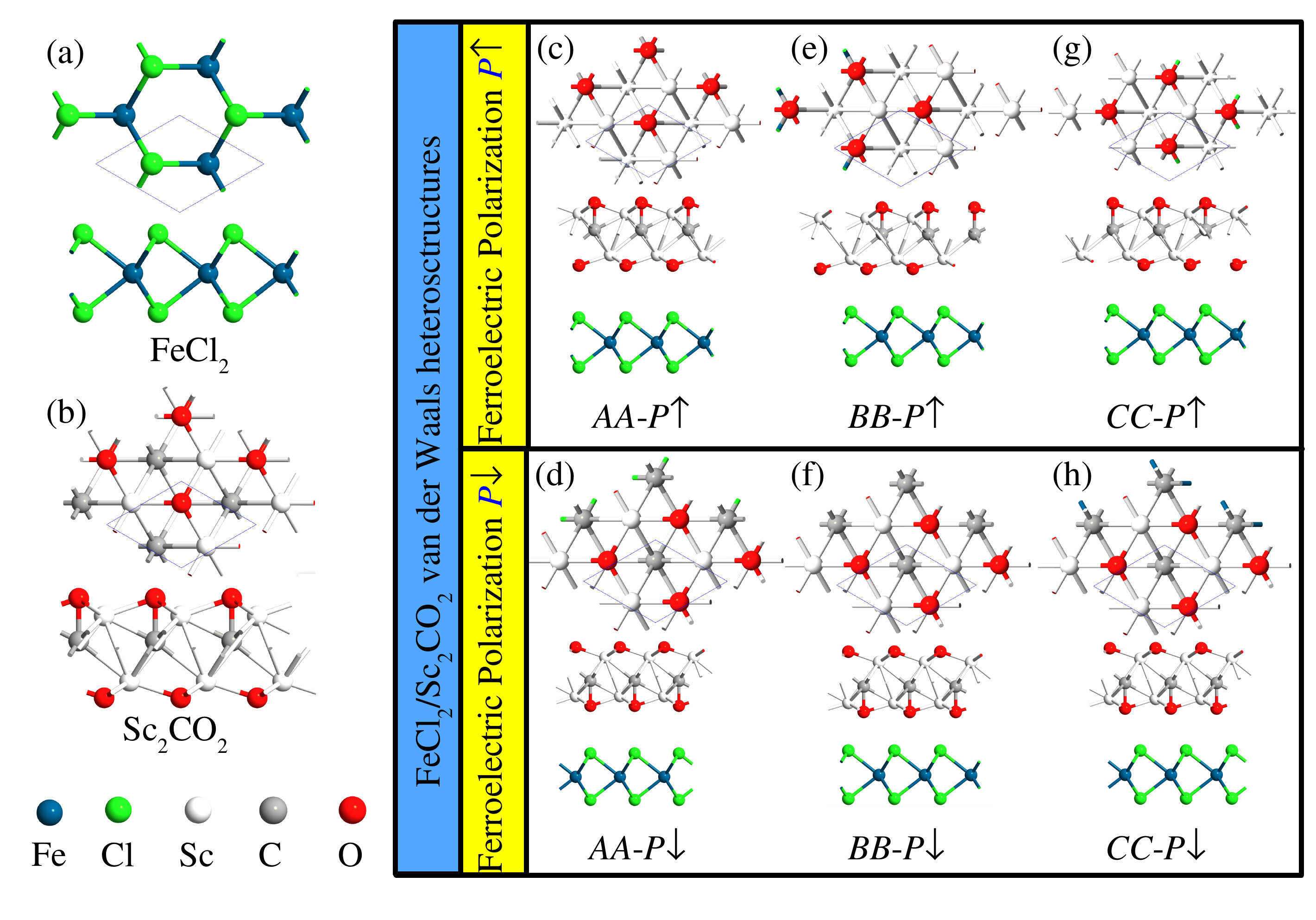}
\caption{Crystal structures of the 2D ferroelectric FeCl$_2$, the 2D ferromagnetic Sc$_2$CO$_2$, and their 2D/2D vdW heterostructures. Top view and side view of the structures of single layer of (a) FeCl$_2$, and (b) Sc$_2$CO$_2$. The blue line represents the unit cell. (c-h) shows the FeCl$_2$/Sc$_2$CO$_2$ vdW heterostructures with different stacking configurations under various polarization states. The $P\uparrow$ and the $P\downarrow$ states are shown in  (c, e, and g) and (d, f, and h) P$\downarrow$, respectively, for $AA$, $BB$, and $CC$ stacking configurations.  The Fe, Cl, Sc, C and O atoms are denoted by dark cyan, green, white, gray and red spheres, respectively. }
	\end{figure*}

\begin{figure*}[t]
\centering
\includegraphics[bb=0 30 800 450, width=7.0 in]{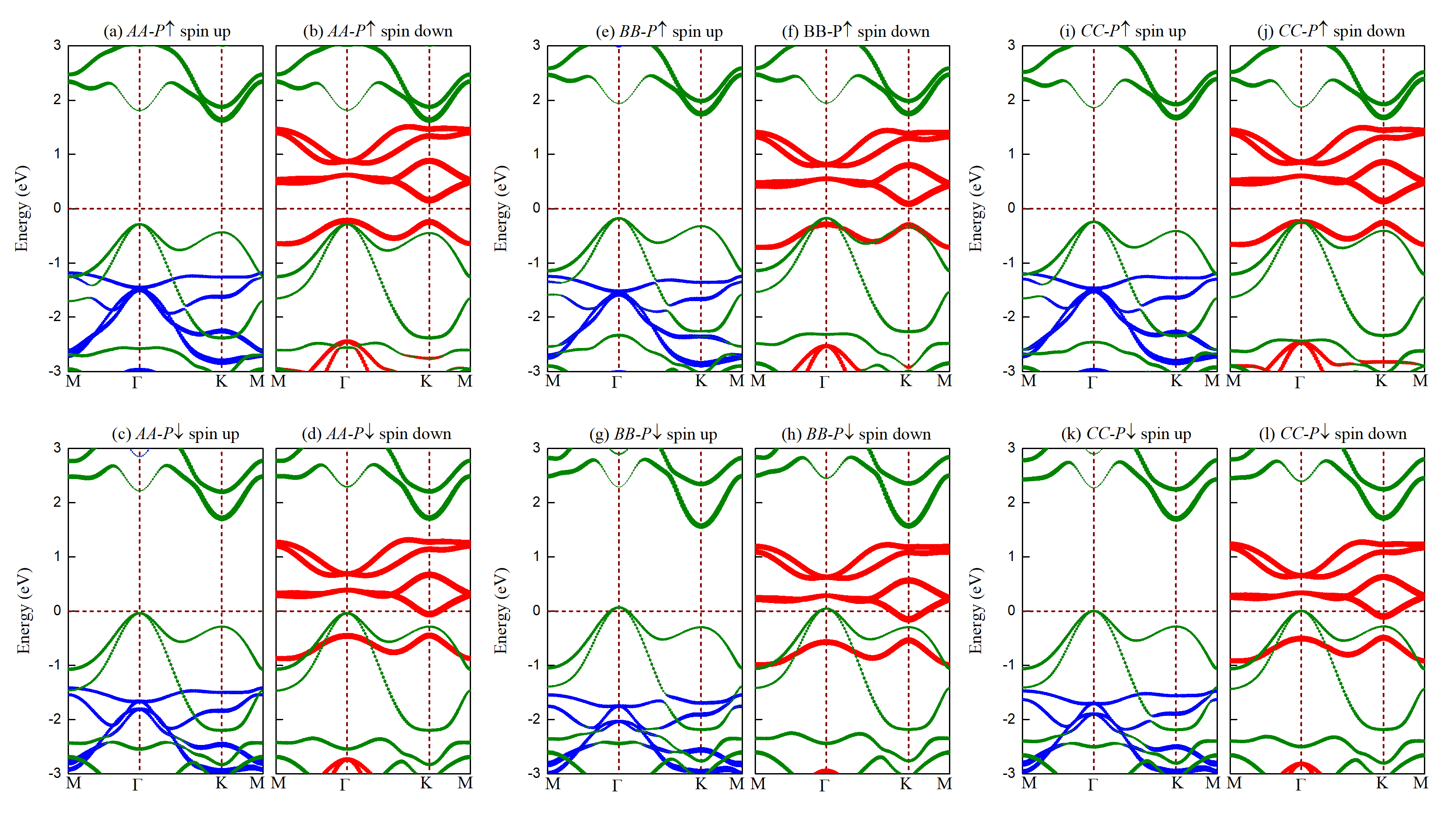}
\caption{ Electronic band structures of (a-d) $AA$-stacked heterostructure, (e-h) $BB$-stacked heterostructure, and (i-l) $CC$-stacked heterostructure under $P\uparrow$ (top panel) and $P\downarrow$ (bottom panel) polarized states. The Fermi level is set to zero. Here green, blue and red symbols denote the contributions from Sc$_2$CO$_2$, spin-up and spin-down electronic states of FeCl$_2$, respectively.}
\end{figure*}	
The union of 2D ferroelectric and ferromagnetic materials leads to the emergence of \emph{multiferroic vdW heterostructures and devices}, in which the all-electric switching of the ferroelectric sub-layer and the magnetic properties of the ferromagnetic sub-layer can be simultaneously combined to render new material properties and device functionalities \cite{Gong2, Lu, Sun, Zhao, Li2}.
Such \emph{magnetoelectric effect} represents a promising route towards the development of ultracompact and multifunctional nanodevices.
The simultaneous presence of two tuning knobs, i.e. the electric fields and the magnetic fields, offers enormous design flexibility in engineering the physical properties of vdW heterostructures for novel electronic, magnetic and spintronic devices with electrically switchable operations.
	
In this work, we investigate the electronic and the transport properties of a multiferroic vdW heterostructures composed of ferromagnetic FeCl$_2$ monolayer and ferroelectric monolayer Sc$_2$CO$_2$ by using first-principle density function theory (DFT) simulations.
Here Sc$_2$CO$_2$ serves as a ferroelectric monolayer with excellent lattice matching with the ferromagnetic FeCl$_2$ monolayer -- highly beneficial for the computational design and simulation of heterostructures and devices.
We show that the electronic band structures of the FeCl$_2$/Sc$_2$CO$_2$ heterostructure can be flexibly tuned between ferromagnetic and half-metallic phase by changing the ferroelectric polarization of Sc$_2$CO$_2$, and such switching is nonvolatile and reversible.
The half-metallic FeCl$_2$/Sc$_2$CO$_2$ heterostructure exhibits a broken-gap Type-III band alignment which can be useful for tunneling-based electronic device applications.
Based on nonequilibrium Green's function (NEGF) transport simulation, we design a proof-of-concept two-terminal nanodevice based on the FeCl$_2$/Sc$_2$CO$_2$ heterostructure and concretely demonstrate the presence of valving effects.
These findings reveals the potential of multiferroic  FeCl$_2$/Sc$_2$CO$_2$ heterostructures as a building block in designing future nanoelectronics and spintronics applications.
Although multiferroic heterostructures are receiving increasing research attention recently, the progress of 2D van der Waals heterostructure multiferroic heterostructures are still in very early stages. In particular, there are very few heterostructures that exhibits reversibly switched from semiconducting to half-metallic behavior by electrically modulating the ferroelectric polarization states. As experimental studies and devices fabrications typically prefers to have larger variety of candidate systems, the results presented in this work offer an important new addition to the multiferroic family that can be useful for future studies.

\section{Computational Methods}
The geometrical optimization and the spin-polarized DFT computations are performed using the Vienna \emph{ab initio} Simulation Package (VASP) \cite{Kresse, Kresse1}, with the generalized gradient approximation (GGA) in the Perdew-Burke-Ernzerhof (PBE) scheme for the exchange-correlation energy functional.
The electron-ion interaction is described by the projector augmented wave (PAW) \cite{Kresse2}.
The kinetic energy cutoff is set to be 500 eV.
The first Brillouin zone integration is sampled with the 15 $\times 15 \times 1$ $k$-point meshes.
	To avoid any artificial interactions between the adjacent slabs, a 20 \AA $ $ vacuum distance is set along the direction vertical to the contact interface.
	Based on the D3 method of Grimme for describing the interlayer vdW interaction \cite{Grimme}, the forces on all atoms less than 0.01 eV/\AA $ $ are achieved for the structural relaxation.
To understand the reliability of DFT-D3 in comparison with other methods, DFT-D2 and vdW-DF are also employed to test the systems studied in this work. Although the interlayer distance varies slightly under different methods, the relative stability of the structures and the calculated band structures are nearly identical to those obtained using DFT-D3 \cite{sm}, thus confirming the reliability of DFT-D3 method.
	The energy convergence accuracy is set to 10$^{-6}$ eV per atom.
	Dipole corrections are included in all calculations.
The results considering the effective Hubbard $U$ (with a semiempirical value of $U=3$ eV \cite{lixr}) and the spin orbit coupling (SOC) are used for comparison, which are presented in Supplementary Materials \cite{sm}.
Finally, the transport properties for a two-probe device systems are calculated using the Atomistix Toolkit (ATK) 2018 package based on the DFT-LCAO and the NEGF method \cite{Soler,Brandbyge,Taylor,Taylor1}.
We use the periodic-type, Neumann-type, and Dirichlet-type boundary conditions \cite{ChengA} in the $x$ (orthogonal to the transport direction), $y$ (perpendicular to the device plane), $z$ (direction of electron transport) directions of the device, respectively. For the $x$-directions, the device is two-dimensionally periodic and infinitely wide. For the $y$-directions, we chose the Neumann boundary condition, which corresponds to a zero electric field at the boundary of the computational box. In the $z$-direction, the boundary condition for the electrostatic potential is determined by the electrostatic potential in the electrodes, corresponding to a Dirichlet boundary condition.

\begin{table*}[t]
\small
\caption{Calculated parameters of the FeCl$_2$/Sc$_2$CO$_2$ multiferroic vdW heterostructures. $d$ is the interlayer distance. $d_{min}$ is the minimum distance between Cl and O atoms. $E_b$ is the binding energy. $\Delta V$ is potential step, defined as $\Delta V = W_{FeCl_2}-W_{Sc_2CO_2}$. $W_{FeCl_2}$ is the work function of the FeCl$_2$.}.
		\begin{tabular*}{\textwidth}{@{\extracolsep{\fill}}lllllllll}
			\hline
			\textbf{Stacking} && \textbf{$AA-P\uparrow$} & \textbf{$AA-P\downarrow$} & \textbf{$BB-P\uparrow$} & \textbf{$BB-P\downarrow$} & \textbf{$CC-P\uparrow$} & \textbf{$CC-P\downarrow$}\\
			\hline
$d$ (\AA)&& 3.248 & 3.119 & 2.725 & 2.46 & 2.856 & 2.677\\
$d_{min}$ (\AA) && 3.248 & 3.119 & 3.35 & 3.13 & 3.45 & 3.31\\
E$_b$ (meV/\AA$^2$) && -75.1 & -82.2 & -82.1 & -95.5 & -80.3 & -89.3\\
Gap (eV) && 0.376 & 0 & 0.264 & 0 & 0.379 & 0\\
$\Delta V$ (eV) && 2.062 & -0.387 & 2.072 & -0.656 & 2.004 & -0.488\\
$W_{FeCl_2}$ (eV) && 5.625 & 5.21 & 5.55&5.371 & 5.613 & 5.307\\
\hline
\end{tabular*}
\end{table*}
	
\section{Results and Discussions}
	
    \begin{figure*}[t]
	\centering
		\includegraphics[bb=0 20 700 300, width=7.0 in]{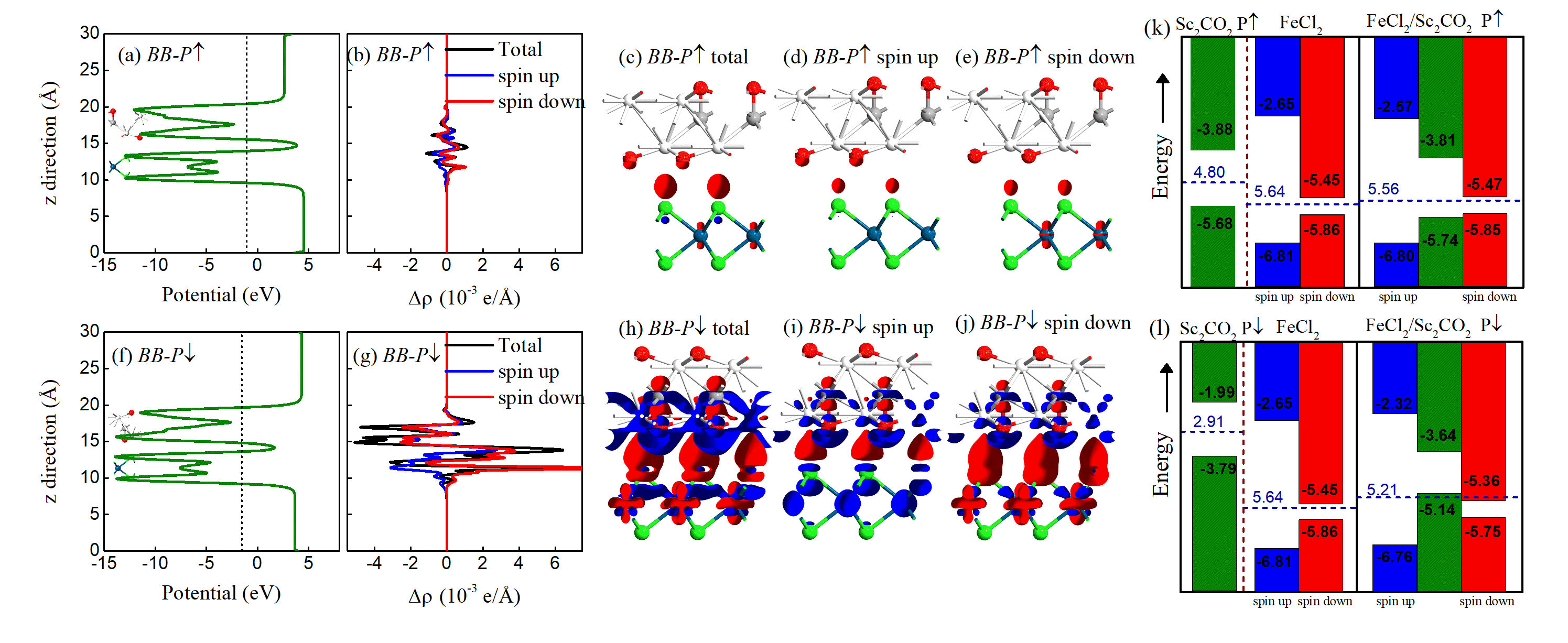}
		\caption{Plane-averaged electrostatic potential of FeCl$_2$/Sc$_2$CO$_2$ (a) $P\uparrow$ and (f) $P\downarrow$ along the Z direction; the plane-averaged differential charge density $\Delta \rho$ of FeCl$_2$/Sc$_2$CO$_2$ (b) $P\uparrow$ and (g) $P\downarrow$; (c-e) and (h-j) the three-dimensional isosurface of the electron density difference (isosurface value is 0.005 e\AA$^{-3}$) of the multiferroic system with $P\uparrow$ and $P\downarrow$ states, respectively, where the blue and red areas represent electron depletion and accumulation, respectively. Band alignments for FeCl$_2$/Sc$_2$CO$_2$ with ferroelectric polarization of (k) $P\uparrow$ and (l) $P\downarrow$, respectively, of isolated layers (left panel) and that after forming heterostructure (right panel). The energies are displayed in the unit of eV. The blue horizontal dashed lines denote the work functions of the systems}
	\end{figure*}

\subsection{Structural and electronic properties}	
The lattice structures of free-standing ferroelectric monolayer FeCl$_2$ and ferromagnetic monolayer Sc$_2$CO$_2$ are 3.366 \AA $ $ and 3.41 \AA, respectively, which agree with the previously reported values \cite{Zhao,Chandrasekaran,Zheng} [see Figs. 1(a) and 1(b)].
The lattice mismatch is less than 1\% when forming the two monolayers stacked vertically to form a 2D/2D vdW multiferroic heterostructure.
Although the lattice optimization of both sub-monolayers can be implemented in the calculation of the 2D/2D heterostructure, such relaxation scheme will lead to additional straining effect that modifies the electronic properties of the heterostructures in addition to the weak van der Waals interaction between the 2D sub-monolayer. The lattice vectors of FeCl2 are thus fixed in our calculations to preserve its physical properties and to minimize strain-induced modifications.
Three typical stacking configurations, denoted $AA$, $BB$, and $CC$, are considered.
Structurally, the three stacking configurations can be described based on the relative location of the O atom: (i) \emph{$AA$ stacking configuration} -- the O atom in contact with FeCl$_2$ is directly under the Cl atom; (ii) \emph{$BB$ stacking configuration} -- the O atom at the contact interface is under an empty space of the FeCl$_2$; and (iii) \emph{$CC$ stacking configuration} -- the O atom is directly under the Fe atom.
	Since the ferroelectric Sc$_2$CO$_2$ has bistable $P$-polarization states, i.e. $P\uparrow$ and $P\downarrow$, six different heterostructures were considered [see Figs. 1(c) to 1(h)].
	Here, when the carbon atom is closer to the top (bottom) layer, the polarization states becomes $P\uparrow$ ($P\downarrow$).
	The detailed structural parameters of the six optimized heterostructures are presented in Table 1.
	The different stacking configurations produces different optimized interlayer distances when the heterostructures is at equilibrium.
	The minimum distance between the atoms at the interface (Cl and O atoms) are much larger than the sum of the covalent radii of Cl and O atoms, thus indicating that vdW interactions are the dominant coupling mechanism between the two monolayers.

	The binding energy, $E_b$, of the FeCl$_2$/Sc$_2$CO$_2$ vdW heterostructure is calculated, which is defined as $E_b = (E_{FeCl_2/Sc_2CO_2} - E_{FeCl_2} - E_{Sc_2CO_2})/A$, where $E_{FeCl_2/Sc_2CO_2}$ is the total energy of FeCl$_2$/Sc$_2$CO$_2$ heterostructure, $E_{FeCl_2}$ and $E_{Sc_2CO_2}$ represent the energy of the isolated monolayer FeCl$_2$ and Sc$_2$CO$_2$, respectively, and $A$ is the surface area at the equilibrium state.
	The binding energy of the combined systems for all stacking modes are summarized in Table 1.
    Many studies have shown that the FeCl2 and Sc2CO2 are stable \cite{Zheng,huh,Chandrasekaran}.
	The calculated binding energies are all negative and are smaller than those of other typical vdW heterostructures \cite{Bjorkman,Padilha}, thus confirming that the proposed 2D/2D heterostructures structures simulated in this work are energetically stable and feasible to explore experimentally.

	The projected band structures of the multiferroic heterostructures are shown in Fig. 2.
	Here green, blue and red symbols denote the contributions from
	Sc$_2$CO$_2$, spin-up and spin-down electronic states of FeCl$_2$, respectively.
	The band structures of isolated Sc$_2$CO$_2$ and FeCl$_2$ are shown in Fig. S1 of the Supplemental Material \cite{sm}.
	In the isolated form, monolayer Sc$_2$CO$_2$ is a semiconductor with a indirect band gap of 1.8 eV \cite{Khazaei,Chandrasekaran}, while monolayer FeCl$_2$ is a ferromagnetic semiconductor with the spin-up and spin-down energy gaps of 4.16 and 0.41 eV, respectively.
	When forming a vdW heterostructures, the electronic band structures of the FeCl$_2$/Sc$_2$CO$_2$ heterostructures exhibit largely a superposition of the electronic states from both materials due to the weak vdW coupling between the two monolayers.
	The retaining of individual electronic bands structures upon forming a vdW heterostructure is a typical features of 2D/2D vdW heterostructures.
	
 \begin{figure*}
		\centering
		\includegraphics[bb=0 60 870 500, width=7.0 in]{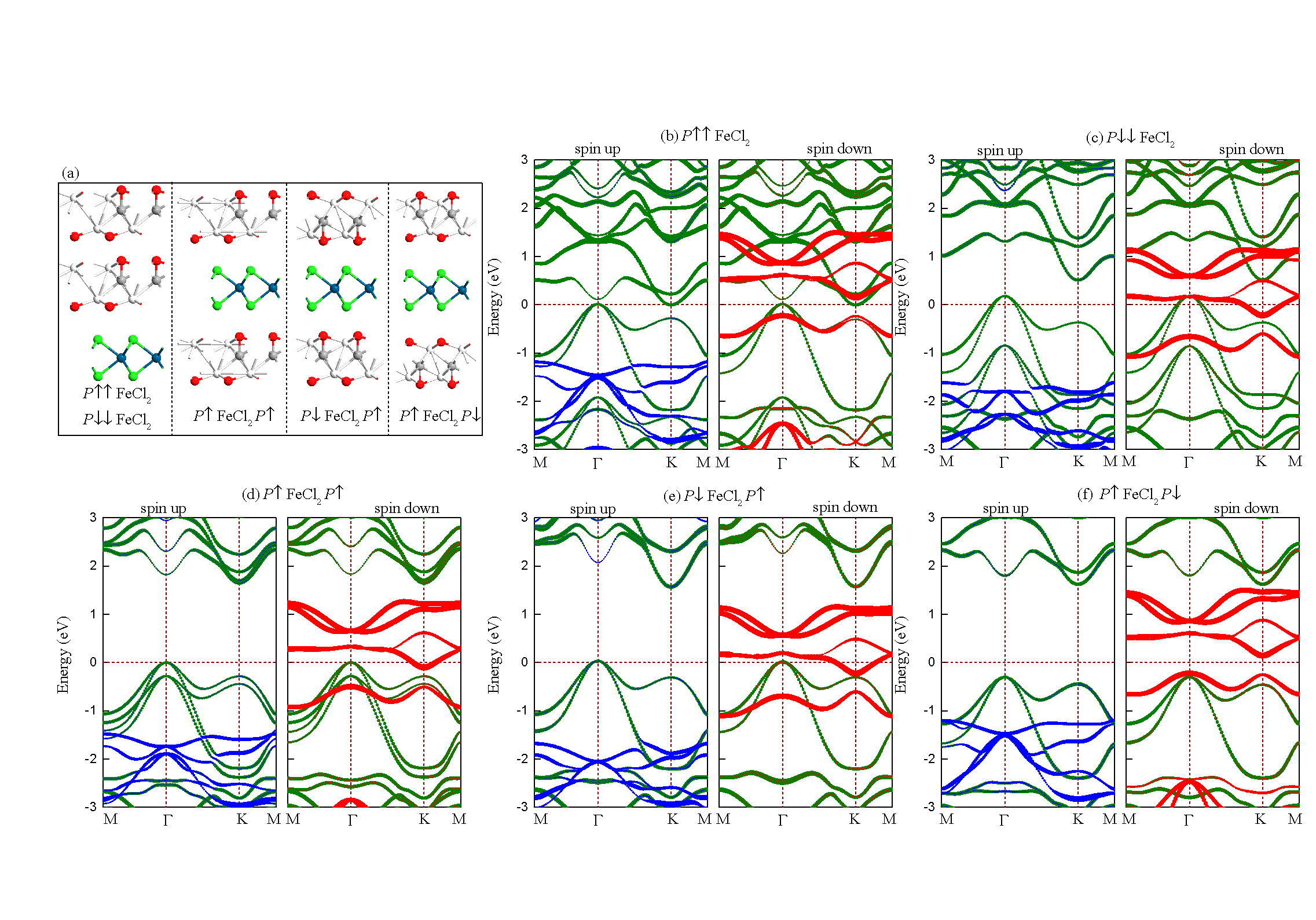}
		\caption{(a) Optimized geometries of four configurations of bilayer Sc$_2$CO$_2$/FeCl$_2$ or Sc$_2$CO$_2$/FeCl$_2$/Sc$_2$CO$_2$ vdW heterostructures; (b-f) electronic band structures of bilayer Sc$_2$CO$_2$/FeCl$_2$ or Sc$_2$CO$_2$/FeCl$_2$/Sc$_2$CO$_2$ vdW heterostructures under different polarized states, respectively.}
 \end{figure*}

	The electronic bands corresponding to the FeCl$_2$ monolayer are sensitively energetically shifted depending on the $P$-polarization states of the Sc$_2$CO$_2$ monolayer.
	In the case of $P\uparrow$-polarized Sc$_2$CO$_2$, the electronic states of Sc$_2$CO$_2$ is up-shifted energetically when compared to that of the independent Sc$_2$CO$_2$ monolayer.
	In this case, a common band gap is retained and the heterostructure behaves as a semiconductor [see Fig. 2, top panel].
	In contrast, the electronic properties is more dramatically modified when the $P$-polarization states of the Sc$_2$CO$_2$ monolayer is switched to the $P\downarrow$-polarization state.
	In this case, the electronic bands of FeCl$_2$ is energetically down-shifted and crosses the Fermi level, thus transforming the FeCl$_2$ into a half-metal.
	Furthermore, the spin-down bands of the FeCl$_2$ monolayer (denoted by red symbols in Fig. 2) overlap with the spin-degenerate valence bands (denoted by green symbols in Fig. 2) of the Sc$_2$CO$_2$ monolayer.
	The heterostructure is thus a \emph{half-metallic semimetal} with zero common band gap when the ferroelectric polarization of the Sc$_2$CO$_2$ monolayer is switched to the $P\downarrow$ state.
	Such contrasting electronic properties are particularly evident in the $BB$ stacking configurations due to its small interlayer distance that strongly couples the two monolayers.
	In terms of heterostructure band alignment, FeCl$_2$/Sc$_2$CO$_2$ in the half-metallic state exhibits a \emph{broken-gap Type-III band alignment} under all stacking configurations, (i.e. when material A and material B merge, the resulting heterostructure is type I if the VBM and CBM are contributed by the same monolayer; is type II if the VBM and CBM are contributed by two different monolayers; and is type III if the VBM and CBM are contributed by two different monolayers and overlap with each other.
Such band alignment type reveals the potential of FeCl$_2$/Sc$_2$CO$_2$ in tunneling field-effect transistor applications.
	
	We further note that when an U$_{eff}$=3 eV is applied to account for a strong correlated effect of the electrons in 3$d$ shell of Fe atoms, the characteristic transformation of the FeCl$_2$ monolayer from the ferromagnetic semiconductor into half-metallic semimetal remains robust (see Fig. S2 of the Supplemental Material \cite{sm}).
	Furthermore, when the SOC is included in the simulations, as shown in Fig. S3 of the Supplemental Material \cite{sm}, the electronic band structures of the heterostructure exhibit no obvious change with only a slight decrease of the energy band gap.
	We further note that the observed transformation of the FeCl$_2$ from semiconducting to semimetallic behaviors are also consistent with other 2D/2D multiferroic vdW heterostructures, such as MnCl$_3$/CuInP$_2$S$_6$ \cite{Li1}, Cr$_2$Ge$_2$Te$_6$/In$_2$Se$_3$ \cite{Gong2}, FeI$_2$/In$_2$Se$_3$ \cite{Sun}, and CrI$_3$/Sc$_2$CO$_2$ \cite{Zhao}.

\subsection{Mechanism of interfacial interaction and charge redistribution}
	
We attribute the switching behavior discussed above to the electrical interactions mediated by the charge transfer across the FeCl$_2$/Sc$_2$CO$_2$ contact interfaces which is appreciably modified between the $P\uparrow$ and the $P\downarrow$ polarization states.
When FeCl$_2$ and Sc$_2$CO$_2$ is constructed into the multiferroic system, the spatial inversion symmetry is broken due to the presence of ferroelectricity, and the charge balance of FeCl$_2$ is perturbed.
As a result, the charges redistribute so to allow the system to reach a new equilibrium.
The Sc$_2$CO$_2$ monolayer with different polarization states exhibit very different charge transfer characteristics when forming the vdW heterostructure as illustrated in Figs. 3(a) to 3(j).
Since the Sc$_2$CO$_2$ is an out-of-plane ferroelectric material, the intrinsic dipole and the in-plane average electrostatic potential differ significantly along the out-of-plane direction for the two $P$-polarization states.
When FeCl$_2$ is interfaced with $P\uparrow$ Sc$_2$CO$_2$, the conduction bands of FeCl$_2$ for spin-up and spin-down subbands are all energetically higher than the valence bands of Sc$_2$CO$_2$, while the valence bands of FeCl$_2$ are lower than that of Sc$_2$CO$_2$.
	Such energy band structures are typical of \emph{type-II band alignment} [Fig. 3(k)] which hinders the electron transfer between FeCl$_2$ and Sc$_2$CO$_2$.
	Nonetheless, the built-in electric field with a direction pointing from the Sc$_2$CO$_2$ to the FeCl$_2$ results in a weak electron transfer.
	Therefore, a small number of electrons accumulates around the FeCl$_2$ as shown in Figs. 3(c) to 3(e).
	Such interfacial charge transfer process enhances the original internal electric field and leads to a potential step of 2.062 eV across the contact interface.
	However, such effect is still insufficient to close the common band gap in the FeCl$_2$/Sc$_2$CO$_2$ heterostructure.

	In contrast, when the polarity state of Sc$_2$CO$_2$ is switched to $P\downarrow$, a larger charge transfer changes occurs as the valence band edge of Sc$_2$CO$_2$ becomes energetically higher than the spin-down conduction band of FeCl$_2$ [Fig. 3(l)].
	The electrons in the valence bands of Sc$_2$CO$_2$ are sufficiently energetic to transit to the spin-up conduction bands of FeCl$_2$.
	As shown in the plane-averaged differential charge density $\Delta \rho$ and the three-dimensional isosurface of the electron density difference of FeCl$_2$/Sc$_2$CO$_2$ heterostructures in Fig. 3(g) and Figs. 3(h) to 3(j), a large amount of charge transfers occurs from Sc$_2$CO$_2$ to FeCl$_2$.
	The partial electrons filling of the spin-down subbands of FeCl$_2$ from the Sc$_2$CO$_2$ valence bands causes the conduction bands to shift energetically downwards towards the Fermi level.
	The overlap between the spin-down conduction band of FeCl$_2$ and the valence band of Sc$_2$CO$_2$ leads to the disappearance of the common gap, and the heterostructure is transformed into a semimetal with Type-III band alignment.

\subsection{Structural and electronic properties of the multiferroic heterostructure with the bilayer ferroelectric materials}

Studies have shown that increasing the number of polar unit cells can make the free carrier density gradually approach the bound charge density. Therefore, the bilayer Sc$_2$CO$_2$ exhibits metallicity due to the strong covalent interaction between carbon and oxygen sublattices facilitates the development of the necessary potential difference\cite{Chandrasekaran}, as shown in Supplemental Material \cite{sm}.
 The potential difference is twice that of a monolayer, which may cause more charge transfer when it contact with the other materials.
 The results show that the nature of electronic structure of FeCl$_2$ to be flexibly switched from semiconductor to half-metallic remains by changing the ferroelectric polarization of bilayer Sc$_2$CO$_2$, as shown in Fig. 4(b) and (c).
 The half-metal phase is more pronounced because of more electron transfer.
 Three more sandwich configurations were constructed, which the monolayer FeCl$_2$ is $BB$-stacked vertically between two monolayer Sc$_2$CO$_2$: $P\uparrow$FeCl$_2$$P\uparrow$, $P\downarrow$FeCl$_2$$P\uparrow$, and $P\uparrow$FeCl$_2$$P\downarrow$, as shown in Fig. 4(a).
 $P\downarrow$FeCl$_2$$P\downarrow$ and $P\uparrow$FeCl$_2$$P\uparrow$ are mirror symmetric, which have the same electronic properties.
 The conduction subband of the spin down of FeCl$_2$ in the $P\downarrow$ polarization state will down-shifted and crosses the Fermi level.
 This property does not change when add another layer of Sc$_2$CO$_2$ with the same polarity, as can be seen from Fig. 4(d).
 However, when a layer of Sc$_2$CO$_2$ with opposite polarity is added, the overall polarity disappears.
 While the internal polarity still exists, and the built-in electric field will redistribute the spatial charge at the interface. Therefore, $P\downarrow$FeCl$_2$$P\uparrow$ heterostructure exhibits half-metal properties in the spin down state, and $P\uparrow$FeCl$_2$$P\downarrow$ heterostructure is semiconducting, which are similar to monolayer ferroelectric materials.

	\subsection{Electrically-switchable transport of a two-channel nanodevice: Quantum transport simulation}
	
	\begin{figure*}[t]
		\centering
		\includegraphics[bb=0 260 820 500, width=7.0 in]{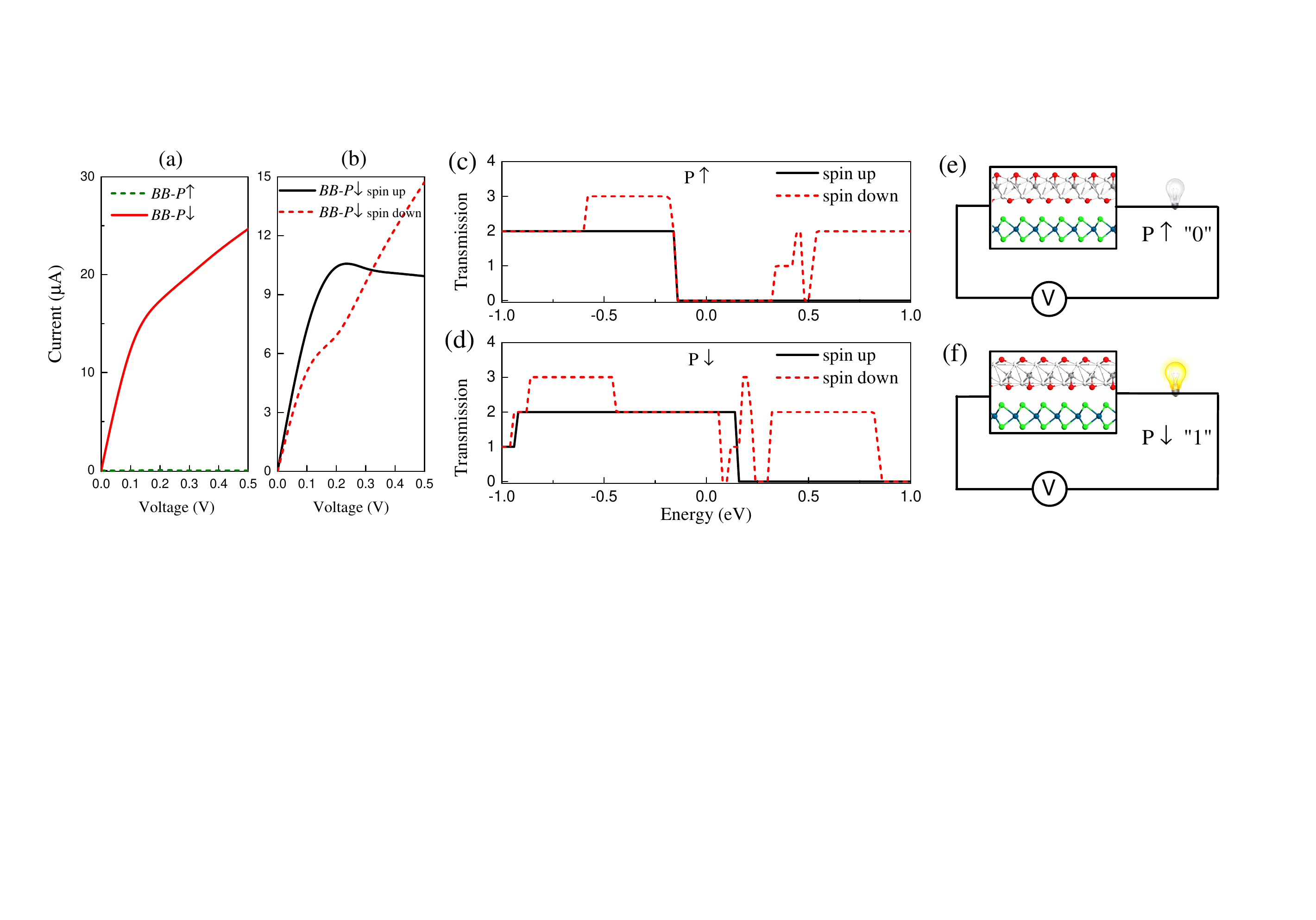}
		\caption{Simulated transport current as a function of the applied bias for FeCl$_2$/Sc$_2$CO$_2$ $BB$-stacking with (a) different ferroelectric polarization states. (b) The spin-up and spin-down currents when the heterostructure is in the $P\downarrow$ state. The spin-dependent transmission spectrum of the device in (c) $P\uparrow$ and (d) $P\downarrow$ polarization states at zero bias. (e) and (f) show the schematic diagrams of a FeCl$_2$/Sc$_2$CO$_2$ multiferroic all-electric-controlled device based on the nonvolatile switching of the $P\uparrow$ and the $P\downarrow$ polarization states of the Sc$_2$CO$_2$ monolayer.}
	\end{figure*}
		
	The two polarization states of $P\uparrow$ and $P\downarrow$ in Sc$_2$CO$_2$ can be reversibly switched by applying an external electric field to the heterostructure, and the polarization state is retained even after the removal of the external electric field.
The energy barrier for switching of polarization from the polar to nonpolar configuration is 0.52 eV per formula unit and proceeds via a multistep process of oxygen atom displacements \cite{Chandrasekaran}. The energy barrier is higher than that of the well-known perovskite ferroelectric LiNbO$_3$(0.26 eV)\cite{yem}.
	The switching of the FeCl$_2$/Sc$_2$CO$_2$ vdW heterostructure between the semiconducting and the half-metalic behavior is thus nonvolatile.
	Importantly, due to the atomic thickness of the FeCl$_2$/Sc$_2$CO$_2$ vdW heterostructure, an external electric field generated by an external gate electrode can readily penetrate across the monolayers, thus allowing such ferroelectric-based polarity switching to be straightforwardly and efficiently achieved all-electrically \cite{Fei}.
	
	Although memory devices based on multiferroic vdW heterostructures have been previously proposed by Zhao et al \cite{Zhao} and Li et al \cite{Li2}, the transport properties and the nanodevice application capabilities of multiferroic vdW heterostructures remain largely unexplored thus far.
	Here we construct a two-probe source-channel-drain \emph{proof-of-concept} device to demonstrate how the vastly contrasting behaviors in $P\uparrow$ and $P\downarrow$ polarization states of the Sc$_2$CO$_2$ monolayer can be harnessed for valving applications.
In the Fig. 5(a), we present the simulated $I-V$ curves at room temperature, where the devices exhibit obvious valving effect with a ON-OFF ratio of 3200. Moreover, the magnitude of the current can reach the order of $\mu$A on the ON state.
The spin-up and spin-down currents exhibit different behaviors with the applied bias voltage in the $P\downarrow$ polarization states [see Fig. 5(b)]. The spin-up current decreases when the external bias exceeds 0.23 V, while the spin-down increases monotonously with the external bias. Such contrasting behaviors arises from the spin-dependent band structures of the heterostructures [see Figs. 2(g) and 2(h)]. The spin-up bands have a large band gap above the Fermi level that suppresses electron transport [Fig. 2(g)]. This is in stark contrasts to the case of spin-down bands in which the energy regime around the Fermi level is populated by the spin-down bands contributed by FeCl2 sublayer [Fig. 2(h)]. Such spin-contrasting band structures directly leads to a contrasting spin-up and spin-down currents as observed in Fig. 5(b).
	We calculate the transmission spectrum of the proposed FeCl$_2$/Sc$_2$CO$_2$ multiferroic vdW heterostructure device at zero bias \cite{caolm,xiacx,Saraiva,Zhao3,caolie, qu_NEGF}.
	The Fermi level is set to the zero-energy location.
	For the P$\uparrow$ state, a large zero transport gap exists around the Fermi level for both spin-up and spin-down channel, which essentially switches off the electrical current in the device [Figs. 5(c) and 5(e)], thus providing a valving effect.
	Such transport gap arises directly from the semiconducting nature of the the FeCl$_2$/Sc$_2$CO$_2$ heterostructure when the Sc$_2$CO$_2$ monolayer is electrically switched to $P\uparrow$-polarization.
	On the other hand, when the Sc$_2$CO$_2$ monolayer is electrically switched to the $P\downarrow$-polarization, the heterostructure becomes semimetallic which opens up the transport channel with a large transmission coefficient around the Fermi level, thus allowing an electrical current to flow in the device [Figs. 5(d) and 5(f)].

\section{Conclusions}

	In summary, we investigated the electronic properties of the FeCl$_2$/Sc$_2$CO$_2$ multiferroic van der Waals heterostructure by using first-principles density functional theory and nonequilirbrium Green's function transport simulations.
	We show that the electronic properties of the FeCl$_2$/Sc$_2$CO$_2$ heterostructures can be reversibly controlled by electrically switching the ferroelectric polarization state of the Sc$_2$co$_2$ monolayer, thus enabling a great wealth of novel functionalities to be derived from FeCl$_2$/Sc$_2$CO$_2$.
	The half-metallic state of FeCl$_2$/Sc$_2$CO$_2$ exhibits Type-III band alignment, thus suggesting a potential in tunneling field-effect transistor applications.
	The quantum transport simulation of a proof-of-concept two-terminal device demonstrated the capability of FeCl$_2$/Sc$_2$CO$_2$ heterostructure for valving applications.
	These findings reveal the potential of FeCl$_2$/Sc$_2$CO$_2$ heterostructure as a building block for designing a next generation of ultracompact and all-electric-controlled spintronics nanodevices.

\section*{Acknowledgements}
L.C. is supported by the National Natural Science Foundation of China (Grant No. 12104136) and by Scientific Research Fund of Hunan Provincial Education Department(Grant No. 21B0622). G. Z. acknowledges the support of National Natural Science Foundation of China (Grant No. 12174100). Y.S.A. acknowledges the support of SUTD Startup Research Grant (Project No. SRT3CI21163). L.K.A. acknowledges the support of Singapore Ministry of Education Academic Research Fund (MOE AcRF) Tier 2 grant (2018-T2-1-007) and A*STAR AME IRG grant (RGAST2007). All the calculations were carried out using the computational resources provided by the National Supercomputing Centre (NSCC) Singapore.



\begin{thebibliography}{99}

\bibitem{Soumyanarayanan} A. Soumyanarayanan, N. Reyren, A. Fert, C. Panagopoulos, \textit{Nature} \textbf{539}, 509 (2006).

\bibitem{Zhang} W. Zhang, P. K. J. Wong, R. Zhu, and A. T. S. Wee, \textit{InfoMat} \textbf{1}, 479 (2019).

\bibitem{liemao} L. M. Cao, G. H. Zhou, Q. Y. Wu, S. Y. Yang, H. Y. Yang, Y. S. Ang, and L. K. Ang, \textit{Phys. Rev. Appl.} \textbf{13}, 054030 (2020).

\bibitem{li} H. Li, S. Ruan, and Y. J. Zeng, \textit{Adv. Mater.} \textbf{31}, 1900065 (2019).

\bibitem{liang} S. J. Liang, B. Cheng, X. Cui, and F. Miao, \textit{Adv. Mater.} \textbf{32}, 1903800 (2019).

\bibitem{liu} Y. Liu, Y. Huang, and X. F. Duan, \textit{Nature} \textbf{567}, 323 (2019).

\bibitem{Gong1} C. Gong and X. Zhang, Two-dimensional magnetic crystals and emergent heterostructure devices, \textit{Science}, \textbf{363}, 706 (2019).

\bibitem{zhong} D. Zhong, K. L. Seyler, X. Linpeng, R. Cheng, N. Sivadas, B. Huang, E. Schmidgall, T. Taniguchi, K. Watanabe, M. A. McGuire, W. Yao, D. Xiao, K. M. C. Fu, and X. Xu, \textit{Sci. Adv.}, \textbf{3} e1603113 (2017).

\bibitem{Novoselov} K. S. Novoselov, A. Mishchenko, A. Carvalho, and A. H. Castro Neto, 2D materials and van der Waals heterostructures, \textit{Science} \textbf{353}, 6298 (2016).

\bibitem{ANG_PRB} Y. S. Ang, L. K. Ang, C. Zhang, and Z. Ma, \textit{Phys. Rev. B} \textbf{93}, 041422 (2016).

\bibitem{ANG_PRB2} Y. S. Ang, S. A. Yang, C. Zhang, Z. Ma, and L. K. Ang, \textit{Phys. Rev. B} \textbf{96}, 245410 (2017).

\bibitem{Cao_APL} L. Cao, G. Zhou, Q. Wang, L. K. Ang, and Y. S. Ang, \textit{Appl. Phys. Lett.} \textbf{118}, 013106 (2021).

\bibitem{Qing_APL} Q. Wu, L. Cao, Y. S. Ang, and L. K. Ang, \textit{Appl. Phys. Lett.} \textbf{118}, 113102 (2021).

\bibitem{Chuong_PRB} C. V. Nguyen, \textit{Phys. Rev. B} \textbf{103}, 115429 (2020).

\bibitem{Chuong_NJC} K. D. Pham, L. V. Tan, M. Idrees, B. Amin, N. N. Hieu, H. V. Phuc, L. T. Hoa, N. V. Chuong, \textit{New J. Chem.} \textbf{44}, 14964-14969 (2020).

\bibitem{Chuong_JPCL} C. Nguyen, N. V. Hoang, H. V. Phuc, Y. S. Ang, C. V. Nguyen, \textit{J. Phys. Chem. Lett.} \textbf{12}, 5076-5084 (2021).

\bibitem{bafekry2020a} A. Bafekry, M. M. Obeid, C. V. Nguyen, M. Ghergherehchi and M. Bagheri Tagani, \textit{J. Mater. Chem. A} \textbf{8}, 13248 (2020).

\bibitem{bafekry2020b} A. Bafekry, C. Stampfl and M. Ghergherehchi, \textit{Nanotechnol.} \textbf{31}, 295202 (2020).

\bibitem{bafekry2020c} A. Bafekry, M. Faraji, A. Abdollahzadeh Ziabari, M. M. Fadlallah, C. V. Nguyen, M. Ghergherehchi and S. A. H. Feghhi, \textit{New J. Chem.} \textbf{45}, 8291 (2021).

\bibitem{bafekry2020d} A. Bafekry and M. Neek-Amal, Phys. Rev. B 101, 085417 (2020).


\bibitem{Shi} H. Shi, H. Pan, Y. W. Zhang, and B. I. Yakobson, \textit{Phys. Rev. B} \textbf{88}, 205305 (2013).

\bibitem{Tongay} S. Tongay, S. S. Varnoosfaderani, B. R. Appleton, J. Wu, and A. F. Hebard, \textit{Appl. Phys. Lett.} \textbf{101}, 123105 (2012).

\bibitem{jiazhang} J. Zhang, J. M. Song, K. P. Loh, J. Yin, J. Ding, M. B. Sullivian, and P. Wu, \textit{Nano Lett.} \textbf{7}, 2370 (2007).

\bibitem{Guo} Y. Guo, B. Wang, X. Zhang, S. Yuan, L. Ma, and J. Wang, \textit{InfoMat} \textbf{2}, 639-655 (2020).

\bibitem{Gong} C. Gong, L. Li, Z. Li, H. Ji, A. Stern, Y. Xia, T. Cao, W. Bao, C. Wang, Y. Wang, Z. Qiu, R. Cava, S. G. Louie, J. Xia, and X. Zhang, \textit{Nature} \textbf{546}, 265 (2017).

\bibitem{Huang} B. Huang, G. Clark, E. Navarro-Moratalla, D. R. Klein, R. Cheng, K. L. Seyler, D. Zhong, E. Schmidgall, M. A. McGuire, D. H. Cobden, W. Yao, D. Xiao, P. Jarillo-Herrero, and X. Xu, \textit{Nature} \textbf{546}, 270 (2017).

\bibitem{song} T. Song, X. Cai, M. W.-Y. Tu, X. Zhang, B. Huang, N. P. Wilson, K. L. Seyler, L. Zhu, T. Taniguchi, K. Watanabe, M. A. McGuire, D. H. Cobden, D. Xiao, W. Yao, and X. Xu, \textit{Science} \textbf{360}, 1214 (2018).

\bibitem{Wang} Z. Wang, L. Guti\'{e}rrez-Lezama, N, Ubrig, M. Kroner, M. Gibertini, T. Taniguchi, K. Watamane, A. Imamo\v{g}lu, E. Giannini, and A. F. Morpurgo, \textit{Nat. Commun.} \textbf{9}, 2516 (2018).

\bibitem{Kim} H. H. Kim, B. Yang, T. Patel, F. Sfigakis, C. Li, S. Tian, H. Lei, and A. W. Tsen, \textit{Nano Lett.} \textbf{18}, 4885 (2018).

\bibitem{Paudel} T. R. Paude, and E. Y. Tsymbal, \textit{ACS Appl. Mater. Interfaces} \textbf{11}, 15781 (2019).

\bibitem{Seyler} K. L. Seyler, D. Zhong, B. Huang, X. Linpeng, N. P. Wilson, T. Taniguchi, K. Watanabe, W. Yao, D. Xiao, M. A. McGuire, K. M. C. Fu, and X. Xu, \textit{Nano Lett.} \textbf{18}, 3823 (2018).

\bibitem{Zhang1} Z. Zhang, X. Ni, H. Huang, L. Hu, and F. Liu, \textit{Phys. Rev. B} \textbf{99}, 115441 (2019).

\bibitem{Cui} C. Cui, F. Xue, W.-J. Hu, and L.-J. Lain, \textit{NPJ 2D Mater. Applications} \textbf{2}, 18 (2018).

\bibitem{Guan} Z. Guan, H. Hu, X. Shen, P. Xiang, N. Zhong, J. Chu, and C. Duan, \textit{Adv. Electron. Mater.} \textbf{6}, 1900818 (2020).

\bibitem{Ding} W. Ding, J. Zhu, Z. Wang, Y. Gao, D. Xiao, Y. Gu, Z. Zhang, and W. Zhu,  \textit{Nat. Commun.} \textbf{8}, 14956 (2017).

\bibitem{Shirodkar} S. N. Shirodkar, and U. V. Waghmare,  \textit{Phys. Rev. Lett. } \textbf{112}, 157601 (2014).

\bibitem{Fei1} R. Fei, W. Kang, and L. Yang, \textit{Phys. Rev. Lett.} \textbf{117}, 097601 (2016).

\bibitem{Wang2} H. Wang, and X. Qian, \textit{Nano Lett.} \textbf{17}, 5027 (2017).

\bibitem{Li1} L. Li, and M. Wu, \textit{ACS Nano} \textbf{11}, 6382 (2017).

\bibitem{Qi} J. Qi, H. Wang, X. Chen, and X. Qian, \textit{Appl. Phys. Lett.} \textbf{113}, 043102 (2018).

\bibitem{Gong2} C. Gong, E. M. Kim, Y. Wang, G. Lee, and X. Zhang, \textit{Nat. Commun.} \textbf{10}, 2657 (2019).

\bibitem{Lu} Y. Lu, R. Fei, X. Lu, L. Zhu, L. Wang, and L. Yang, \textit{ACS Appl. Mater. Interfaces} \textbf{12}, 6243 (2020).

\bibitem{Sun} W. Sun, W. Wang, D. Chen, Z. Cheng, and Y. Wang, \textit{Nanoscale} \textbf{11}, 9931 (2019).

\bibitem{Zhao} Y. Zhao, J.-J. Zhang, S. Yuan, and Z. Chen, \textit{Adv. Funct. Mater.} \textbf{29}, 1901420 (2019).

\bibitem{Li2} Z. Li, and B. Zhou, \textit{J. Mater. Chem. C} \textbf{8} , 4534-4541 (2020).



\bibitem{Kresse} G. Kresse and J. Furthm\"{u}ller, \textit{Phys. Rev. B} \textbf{54}, 11169 (1996).

\bibitem{Kresse1} G. Kresse and J. Furthm\"{u}ller, \textit{Comput. Mater. Sci.} \textbf{6}, 15 (1996).

\bibitem{Kresse2} G. Kresse and D. Joubert, \textit{Phys. Rev. B} \textbf{59}, 1758 (1999).

\bibitem{Grimme} S. Grimme, J. Antony, S. Ehrlich and H. Krieg, \textit{J. Chem. Phys.} \textbf{132}, 154104 (2010).

\bibitem{sm} See Supplemental Material at XXXXX for more details on the band structures of isolated monolayers, projected band structures of the heterostructures with SOC, LDA+$U$ ($U_{eff} = 3$ eV) and different vdW correction methods.
    
\bibitem{lixr} X. Li, Z. Zhang and H. Zhang, \textit{Nanoscale Adv} \textbf{2}, 495-501 (2020).


\bibitem{Soler} J. M. Soler, E. Artacho, J. D. Gale, A. Garc\'{\i}a, J. Junquera, P. Ordej\'{o}n, and D. S\'{a}chez-Portal, \textit{J. Phys. Condens. Matter} \textbf{14}, 2745 (2002).

\bibitem{Brandbyge} M. Brandbyge, J. L. Mozos, P. Ordej\'{o}n, J. Taylor, and K. Stokbro, \textit{Phys. Rev. B} \textbf{65}, 165401 (2002).

\bibitem{Taylor} J. Taylor, H. Guo, and J. Wang, \textit{Phys. Rev. B} \textbf{63}, 245407 (2001).

\bibitem{Taylor1} J. Taylor, H. Guo, and J. Wang, \textit{Phys. Rev. B} \textbf{63}, 121104 (2001).

\bibitem{ChengA} A. H. D. Cheng, D. T. Cheng, \textit{Eng. Anal. Bound. Elem.} \textbf{29}, 268 (2005).
    
\bibitem{Chandrasekaran} A. Chandrasekaran, A. Mishra, and A. K. Singh, \textit{Nano Lett.} \textbf{17}, 5, 3290 (2017).

\bibitem{Zheng} H. Zheng, H. Han, J. Zheng, and Y. Yan, \textit{Solid State Commun.} \textbf{271}, 66 (2018).

\bibitem{huh} H. Hu, W.-Y. Tong, Y.-H. Shen, X. Wang, C.-G. Duan, \textit{npj Comput. Mater.} \textbf{6}, 129 (2020).

\bibitem{Bjorkman} T. Bj\"{o}rkman, A. V. Krasheninnikov and R. M. Nieminen, \textit{Phys. Rev. Lett.} \textbf{108}, 235502 (2012).

\bibitem{Padilha} J. F. Padilha, A. Fazzio, and Ant\^{o}nio J. R. da Silva, \textit{Phys. Rev. Lett.} \textbf{114}, 066803 (2015).


\bibitem{Khazaei} M. Khazaei, M. Arai, T. Sasaki, C.-Y. Chung, N. S. Venkataramanan, M. Estili, Y. Sakka, Y. Kawazoe, \textit{Adv. Funct. Mater.} \textbf{23}, 2185 (2013).

\bibitem{yem} M. Ye, D. Vanderbilt, \textit{Phys. Rev. B: Condens. Matter Mater. Phys.} \textbf{93}, 134303 (2016).

\bibitem{Fei} Z. Fei, W. Zhao, T. A. Palomaki, B. Sun, M. K. Miller, Z. Zhao, J. Yan, X. Xu, and D. H. Cobden, \textit{Nature} \textbf{560}, 336 (2018).

\bibitem{xiacx} C. Xia, J. Du, M. Li, X. Li, X. Zhao, T. Wang, and J. Li, \textit{Phys. Rev. Appl.} \textbf{10}, 054064 (2018).

\bibitem{caolie} L. M. Cao, X. B. Li, M. Zuo, C. X. Jia, W. H. Liao, M. Q. Long, and G. H. Zhou, \textit{J. Magn. Magn. Mater.} \textbf{485}, 136 (2019).

\bibitem{qu_NEGF} H. Qu, W. Zhou, S. Guo, Z. Li, Y. Wang, and S. Zhang, \textit{Adv. Electron. Mater.} \textbf{5}, 1900813 (2019).

\bibitem{Saraiva} A. Saraiva-Souza, M. Smeu, J. G. da Silva Filho, E. C. Gir\~{a}o, and H. Guo, \textit{J. Mater. Chem. C} \textbf{5}, 11856 (2017)).

\bibitem{Zhao3} C. C. Zhao, S. H. Tan, Y. H. Zhou, R. J. Wang, X. J. Wang and K. Q. Chen, \textit{Carbon} \textbf{113}, 170 (2017).

\bibitem{caolm} L. Cao, Y. S. Ang, Q. Wu, and L. K. Ang, \textit{Phys. Rev. B} \textbf{101}, 035422 (2020).

\end{thebibliography}
\end{document}